\documentclass[epj]{svjour}

\usepackage{graphicx}
\usepackage{fancyhdr}
\def\makeheadbox{{
 }}
\def\mytitle{My title} 
\def\myauthors{My name}  
\def\mytype{My type of session}
\def\mysession{My session}

\def\mytitle{Gravitino Dark Matter with Broken R-parity} 
\def\myauthors{Alejandro Ibarra}    
\def\mytype{Contributed Talk}    
\def\mysession{Cosmology and Astrophysics}

\def\lsim{\mathrel{\raise.3ex\hbox{$<$\kern-.75em\lower1ex\hbox{$\sim$}}}}
\def\gsim{\mathrel{\raise.3ex\hbox{$>$\kern-.75em\lower1ex\hbox{$\sim$}}}}
\newcommand{\Rp}{\mbox{$\not \hspace{-0.15cm} R_p$}}

\begin{document}
\title{Gravitino Dark Matter with Broken R-parity}
\author{Alejandro Ibarra\inst{1}
\thanks{\emph{Email:} alejandro.ibarra@desy.de}%
}                     
%
%
\institute{Deutsches Elektronen-Synchrotron DESY, Hamburg, Germany}
%
\date{}
\abstract{Scenarios with gravitino dark matter face
potential cosmological problems induced by the 
presence of the Next-to-Lightest Supersymmetric Particle (NLSP) 
at the time of Big Bang Nucleosynthesis (BBN). A very simple,
albeit radical, solution to avoid all these problems consists 
on assuming that $R$-parity is slightly violated, since in this
case the NLSP decays well before the onset of BBN. Remarkably,
even if the gravitino is no longer stable in this scenario, 
it still constitutes a very promising dark 
matter candidate. In this talk we review the motivations for this
scenario and we present a model that 
naturally generates $R$-parity breaking parameters of the 
right size to produce a consistent thermal history 
of the Universe, while satisfying at the same time all the 
laboratory and cosmological bounds on $R$-parity breaking.
We also discuss possible signatures of this scenario at gamma
ray observatories and at colliders.
} 

\maketitle
\begin{picture}(0,0)
\put(400,300){DESY 07-184}
\end{picture}
\vspace{-1.3cm}

\section{Motivation}

The gravitino, when it is the lightest supersymmetric particle,
constitutes a very promising candidate for the dark matter
of the Universe. The interactions of the gravitino 
are completely fixed by the symmetries, and the thermal relic 
density is calculable in terms of very few parameters, 
the result being~\cite{Bolz:2000fu}
\begin{equation}
\Omega_{3/2} h^2\simeq 0.27
    \left(\frac{T_R}{10^{10}\,{\rm GeV}}\right)
    \left(\frac{100 \,{\rm GeV}}{m_{3/2}}\right)
    \left(\frac{m_{\widetilde g}}{1\,{\rm TeV}}\right)^2\;,
\label{relic-abundance}
\end{equation}
while the relic density inferred by WMAP for the $\Lambda$CDM model is
$\Omega_{3/2} h^2\simeq 0.1$~\cite{Spergel:2006hy}.
In this formula, $T_R$ is the reheating temperature of the Universe,
$m_{3/2}$ is the gravitino mass and $m_{\widetilde g}$ is the 
gluino mass. It is indeed remarkable that the correct relic density
can be obtained for typical supersymmetric parameters, 
$m_{3/2}\sim 100\,{\rm GeV}$, $m_{\widetilde g}\sim 1\,{\rm TeV}$,
and a high reheating temperature, $T_R\sim 10^{10}\,{\rm GeV}$, as
required by baryogenesis through the mechanism of 
thermal leptogenesis~\cite{Fukugita:1986hr}.

If thermal leptogenesis is the correct mechanism to
generate the observed baryon asymmetry, then 
the NLSP can pose serious cosmological difficulties. 
If one assumes exact $R$-parity conservation,
as is commonly done in the literature, then the NLSP can only
decay into gravitinos and Standard Model particles with a decay
rate strongly suppressed by the Planck mass, yielding a lifetime
\begin{equation}
\label{lifetime-NLSP}
\tau_{\rm NLSP}\simeq 9~{\rm days}
\left(\frac{m_{3/2}}{10{\rm GeV}}\right)^2
\left(\frac{150{\rm GeV}}{m_{\rm NLSP}}\right)^5\;.
\end{equation}

The high reheat temperature for the Universe required by thermal 
leptogenesis, $T_R\gsim 10^9$ GeV~\cite{bound}, and the 
requirement of correct relic abundance, Eq.~(\ref{relic-abundance}), imply
$m_{3/2}\gsim 5$ GeV for a gluino mass of $m_{\widetilde g}=500$GeV.
Therefore, the lifetime of the NLSP is typically of several
days and it is present at the time of BBN,
potentially jeopardizing the successful predictions of the standard
scenario. Indeed, this is the case for most supersymmetric scenarios
with gravitino dark matter. Namely, when the NLSP is a neutralino, 
its late decays into hadrons can dissociate the primordial 
elements~\cite{Kawasaki:2004qu}, and when the NLSP is a stau, 
its presence during BBN catalyzes the production of $^6$Li, 
yielding a primordial abundance in stark conflict with 
observations~\cite{Pospelov:2006sc}. 

One simple, albeit radical, solution to the problems induced
by the NLSP is to assume that $R$-parity is not exactly 
conserved~\cite{Buchmuller:2007ui}. 
In fact, there is no deep theoretical reason why $R$-parity should be 
exactly conserved, and although present experiments indicate that 
$R$-parity is approximately conserved, a slight violation of it cannot 
be precluded. Without imposing any {\it ad hoc} discrete symmetry, 
the superpotential of the Minimal Supersymmetric Standard 
Model (MSSM) reads~\cite{Barbier:2004ez}
\begin{eqnarray}
\label{Rp-superpotential}
W&=&W_{Rp} + \frac{1}{2} \lambda_{ijk} L_i L_j e^c_k
+ \lambda'_{ijk}\, L_i Q_j d^c_k \nonumber \\
&&+ \frac{1}{2} \lambda''_{ijk} u^c_i d^c_j d^c_k + \mu_i L_i H_u\;,
\end{eqnarray}
where $W_{Rp}$ is the familiar superpotential with $R$-parity
conserved. 

When $R$-parity is broken, new decay channels are open
for the NLSP, apart from the strongly suppressed gravitational
decay into gravitinos and Standard Model particles. Namely,
when the NLSP is mainly a right-handed stau, it can decay
$\widetilde \tau_R \rightarrow \mu\, \nu_{\tau}$, through the
coupling $\lambda_{323}$, with lifetime
\begin{equation}
\tau_{\widetilde \tau} \simeq 10^3 {\rm s} 
\left(\frac{\lambda_{323}}{10^{-14}}\right)^{-2}
\left(\frac{m_{\widetilde \tau}}{{100~{\rm GeV}}}\right)^{-1}.
\end{equation}
Thus, even with a tiny amount of $R$-parity violation,  
$\lambda\gsim 10^{-14}$, the stau NLSP will decay before 
it can have any significant 
impact on Big Bang Nucleosynthesis. A similar argument can be applied
for the case of neutralino NLSP with analogous conclusions.

On the other hand, there are very stringent laboratory and 
cosmological constraints on the size of the $R$-parity breaking 
parameters. The strongest bound for the heaviest generation 
stems from the requirement of successful baryogenesis. Since
$\lambda$, $\lambda'$  and $\lambda''$ violate lepton and 
baryon number, if they are on thermal equilibrium at the same
time as the sphaleron processes, a preexisting baryon asymmetry 
will be erased. Therefore, the requirement that these couplings
are out of thermal equilibrium before the electroweak phase
transition translates into the bound~\cite{cxx91}
\begin{equation}
\lambda, \lambda',\lambda'' \lsim 10^{-7}\;.
\end{equation}
This bound is a sufficient but not a necessary condition, 
and could be relaxed for some flavour structures.

Interestingly, if the size of the $R$-parity violating
couplings is in this range, $10^{-14}\lsim \lambda, \lambda' \lsim 10^{-7}$,
the gravitino still constitutes a viable dark matter 
candidate~\cite{Takayama:2000uz}. The decay rate is doubly 
suppressed by the inverse Planck mass and by the $R$-parity 
breaking couplings, yielding a lifetime
\begin{equation}
\label{grav-lifetime}
\tau_{3/2}\ \sim \ 10^{26} {\rm s}\  
\left(\frac{\lambda}{10^{-7}}\right)^{-2}
\left(\frac{m_{3/2}}{10~{\rm GeV}}\right)^{-3}\;,
\end{equation}
which is several orders of magnitude longer than the age of the
Universe.

In the next sections we will present a model that can
naturally accommodate our favoured range of $R$-parity breaking 
parameters, $10^{-14}\lsim \lambda, \lambda' \lsim 10^{-7}$,
and we will discuss the possible experimental signatures of 
this scenario, both at gamma ray observatories and at colliders.

\section{A model for small (and peculiar) $R$-parity violation}
\label{sec:model}

We will use for convenience $SO(10)$ notation, although
our model is not necessarily embedded into a Grand Unified group.
Then, quarks and leptons will be denoted by ${\bf 16_i}$ and the 
Higgses by ${\bf 10}_H$. In order to give Majorana masses to
neutrinos $B-L$ has to be broken, either by the vacuum expectation value of 
a ${\bf \overline{16}}$, ${\bf  16}$ (with $B-L=\pm 1$) or of 
a ${\bf  126}$ (with $B-L=2$). 
To have only particles with small representations we will break 
$B-L$ with a ${\bf \overline{16}}$ and a ${\bf  16}$. Thus,
$R$-parity will be necessarily violated after the spontaneous
breaking of the $B-L$ symmetry at the scale $v_{B-L}$.

With this matter content there are two kind of terms in the
superpotential allowed by the $SO(10)$ symmetry. 
The terms ${\bf 16_i 16_j 10_H}$ and 
$\frac{1}{M_P}{\bf 16_i 16_j {\overline{16}}\ {\overline{16}}}$ 
are ``good terms'',
since they produce Dirac masses and a Majorana mass term
for the right-handed neutrinos, respectively. On the
contrary, the terms ${\bf 16_i 16 10_H}$ and 
$\frac{1}{M_P}{\bf 16_i 16_j 16_k 16}$ are ``bad terms''; the first
one produces after $B-L$ breaking the $R$-parity
violating bilinear term $v_{B-L} L H_u$ that in turn produces
too large neutrino masses, while the
second one produces the $R$-parity violating trilinear terms
$\frac{v_{B-L}}{M_P}u^c d^c d^c$, $\frac{v_{B-L}}{M_P}Q L d^c$
that induce too rapid proton decay. It is interesting to note
that both ``bad terms'' are generated when the ${\bf 16}$
representation acquires a large vacuum expectation value. 
Since the existence of the ${\bf 16}$ representation in the
spectrum is unavoidable, the only way to suppress these 
couplings is by means of additional symmetries.

To this end, we will impose in our model a $U(1)_R$ symmetry
and the following assignment of charges:
\begin{center}
\begin{tabular}{|c|cccc|c|}
\hline 
  &${\bf 16}_i$ & ${\bf 10}_H$ & ${\bf \overline {16}}$& ${\bf 16}$& 
${\bf 1}$\\
\hline 
$R$ & 1 & 0 & 0 &-2 & -1\\
\hline
\end{tabular}
\end{center}
where the singlet ${\bf 1}$ has been introduced in order to break the
$R$-symmetry. With this assignment of charges, holomorphicity
guarantees that after $B-L$ breaking no $R$-parity violating
term will be generated from the superpotential at any order in 
perturbation theory.
Nevertheless, the K\"ahler potential is not holomorphic
and could be a source of $R$-parity violation. Indeed, terms such as 
${\bf 1}\, {\bf \overline{16}}^\dagger 
{\bf 16}_i {\bf 10}_H$, ${\bf 1}^\dagger {\bf\overline{16}}\,
{\bf 16}_i {\bf 10}_H$ can appear in the K\"ahler potential,
and will eventually produce bilinear $R$-parity violation.

With these elements in mind, it is relatively simple to construct 
a model that produces small lepton number violation and tiny baryon 
number violation. In Standard Model notation, the previous
model is:
\begin{center}
\begin{tabular}{|c|cccc|ccc|}
\hline 
  & $Q$, $u^c$, $e^c$ & $H_u$, $H_d$ 
& $N$ & $N^c$ & $\Phi$ & $Z$ & $X$ \\
& $d^c$, $L$,  $\nu^c$ &  & & & & & \\
\hline 
$B-L$& $\pm 1/3,\pm 1$  &  0 &1 & -1 &0 & 0 & 0 \\
\hline
$R$ & 1 & 0 & 0 &-2 & -1& 0 & 4\\
\hline
\end{tabular}
\end{center}
where the spectator fields $\Phi$, $Z$ and $X$ 
have been introduced in order to break  $B-L$ (and $R$-parity), 
to break supersymmetry, and to 
ensure that $\langle N \rangle = \langle N^c \rangle = 
v_{B-L}$, respectively. After these singlets acquire a vacuum 
expectation value, the effective superpotential reads
$W\simeq W_{R_p}+ W_{\nu^c} +W_{\Rp}$.
Here, $W_{R_p}$ is the $R$-parity conserving superpotential
of the MSSM and $W_{\nu^c}$ is the part of the superpotential
involving right-handed neutrinos. In this model the heaviest right-handed 
neutrino mass is given by $M_3\sim v^2_{B-L}/M_P$, which is naturally large,
leading through the see-saw mechanism to small neutrino masses.
Finally, $W_{\Rp}$ is the $R$-parity breaking superpotential given in 
Eq.(\ref{Rp-superpotential}). We will choose to work in the 
$L_i-H_d$ basis where $\mu_i=0$  and thus all the $R$-parity violation
is encoded in the trilinear terms. In this basis,
\begin{equation}
\lambda\sim C \frac{v_{B-L}^2}{M_P^2} h^e\;,~
\lambda'\sim C \frac{v_{B-L}^2}{M_P^2} h^d\;,~
\lambda''\sim m_{3/2} \frac{v_{B-L}^4}{M_P^5}\;. 
\end{equation}

The lepton number violating couplings are suppressed compared
to the charged lepton and down-type quark Yukawa couplings,
$h^e$ and $h^d$ respectively,  by
$v_{B-L}^2/M_P^2$ and by some coefficients $C=1.0...0.01$
that depend on the flavour structure of the K\"ahler
potential. On the other hand, the baryon number violating
coupling only arises after supersymmetry breaking and is
suppressed by higher powers of the Planck mass.
To estimate the scale of $B-L$ breaking and the size
of the coefficients $C$, we will use the flavour model proposed
in~\cite{by99}. For this particular model, we obtain 
$\lambda_{3ij}, \lambda'_{3ij}\sim 10^{-8}$, within the
desired range $10^{-14}\lsim \lambda, \lambda' \lsim 10^{-7}$,
and $\lambda''\sim 10^{-28}$, yielding negligible rates 
for proton decay~\cite{Buchmuller:2007ui}.

\section{Signatures at gamma-ray observatories}

When $R$-parity is broken, gravitino decays could be happening
at fast enough rates to make its decay products detectable in future 
experiments. In particular, the photon flux produced by the gravitino
decay could be detected as an extragalactic diffuse gamma-ray flux 
with a characteristic spectrum, that is originated from the decay
of gravitinos at cosmological distances and from the decay of 
gravitinos in the Milky Way halo.

In the case that the gravitino is lighter than the $W$ boson,
the gravitino decays mainly into a photon and a neutrino, producing
a photon spectrum consisting in a monochromatic line. If 
gravitinos constitute the dominant component of dark matter,
the decay of gravitinos at cosmological distances will be detected
at the Earth as a perfectly isotropic diffuse gamma ray background,
with an energy spectrum corresponding to a red-shifted monochromatic
line:
\begin{equation}
\label{photon-flux}
 E^2 {dJ_{eg}\over dE} =
 C_\gamma 
\left[1+\frac{\Omega_{\Lambda}}{\Omega_{M}}y^3\right]^{-1/2}
y^{5/2}\ \theta\left(1 - y\right)\; , 
\end{equation}
where $y=2E/m_{3/2}$ and 
\begin{equation}
C_\gamma = \frac{ \Omega_{3/2} \rho_c}{8\pi\tau_{3/2}H_0 \Omega_M^{1/2}}
=  \frac{10^{-6}\ \rm GeV}{{\rm cm}^2\ {\rm str}~{\rm s}}
\left(\frac{\tau_{3/2}}{ 10^{27}\ {\rm s}} \right)^{-1}\;.
\end{equation}
Here, $\Omega_{3/2} h^2 =0.1$, 
$\rho_c = 1.05\; h^2 \times 10^{-5} {\rm GeV} {\rm cm}^{-3}$,
$\Omega_M = 0.25$,
$H_0 = h\; 100\; {\rm km}\; {\rm s}^{-1}\; {\rm Mpc}^{-1}$ 
with $h=0.73$~\cite{PDG} and $ \tau_{3/2} $ 
is given by Eq.~(\ref{grav-lifetime}).

In addition to the contribution to the photon flux 
from the decay of gravitinos at cosmological distances,
one also expects a monochromatic line stemming from the decay
of gravitinos in the Milky Way halo. This contribution reads:
\begin{equation}
\label{photon-flux-halo}
 E^2 {dJ_{halo}\over dE} = D_\gamma\; 
\delta\left(1 - \frac{2E}{m_{3/2}}\right)\; , 
\end{equation}
where
\begin{equation}
D_\gamma = \frac{1}{8\pi\tau_{3/2}}
\int_{l.o.s.} \rho_{halo} (\vec{l}) d\vec{l} \; .
\end{equation}
Since the Earth is not located at the center
of the Galaxy, the gamma-ray flux from the decay of gravitinos
in the halo is slightly anisotropic. 
For typical halo models,
we find that the halo component dominates over the cosmological
one, giving rise to a slightly anisotropic gamma ray flux with
an energy spectrum dominated by a monochromatic line.

There are very stringent limits on the photon flux produced
by decaying dark matter from the observations of the Energetic
Gamma Ray Experiment Telescope (EGRET) aboard the Compton Gamma
Ray Observatory. Assuming that one understands the galactic foreground,
one can extract from the EGRET data the extragalactic diffuse component.
The first analysis by Sreekumar {\it et al.}~\cite{egret} 
gave a roughly isotropic extragalactic flux with an energy spectrum
described by the power law
\begin{equation}
\label{Sreekumar}
E^2 \frac{dJ}{dE} = 1.37 \times 10^{-6}\ 
\left(\frac{E}{1~\rm{GeV}}\right)^{-0.1} 
\frac{ \rm GeV}{{\rm cm}^2\ {\rm str}~{\rm s}}
\end{equation}
in the energy range 50 MeV--10 GeV. The improved analysis
of the galactic foreground by Strong {\it et al.}~\cite{smr05},
optimized in order to reproduce the galactic emission,
shows a power law behavior between  50 MeV--2 GeV, but
a clear excess between 2--10 GeV, roughly the same
energy range where one would expect a signal from gravitino
decay. In view of all the systematic uncertainties involved 
in the extraction of the signal from the galactic foreground, 
we find it premature to attribute this excess to the gravitino
decay. We find nevertheless this coincidence as
interesting and deserving further attention.
The  upcoming satellite-based gamma ray experiments GLAST and 
AMS-02 will measure the energy spectrum with unprecedented accuracy,
providing very valuable information for the scenario of decaying 
gravitino dark matter. 

\begin{figure}
\includegraphics[width=0.45\textwidth,angle=0]{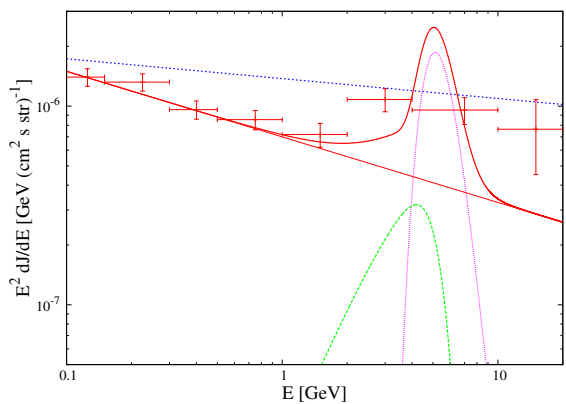}
\caption{Energy spectrum of extragalactic (long-dashed green)
and halo signal (dotted magenta)  compared to
the EGRET data, for $ \tau_{3/2} = 10^{27}\ {\rm s} $ 
and $ m_{3/2} = 10\ {\rm GeV} $. 
The data points are the EGRET extragalactic background
extracted by Strong {\it et al.} in \cite{smr05}, while
the short-dashed (blue) line shows the powerlaw fit 
from Sreekumar {\it et al.}, Eq.~(\ref{Sreekumar}).
Finally, the solid (thick red) line shows the sum of these contributions
 with a powerlaw background (thin red line), which has
been obtained fitting the low energy EGRET points.}
\label{fig:1}       
\end{figure}

We show in Fig.\ref{fig:1} the expected signal for a decaying gravitino
with a mass of 10 GeV and a lifetime of $10^{27}$ s~\cite{Bertone:2007aw}.
To compare our results with 
the EGRET data~\cite{smr05}, also shown in the figure, we have averaged
the halo signal over the whole sky excluding a band of $\pm 10^\circ$
around the Galactic disk, and we have used an energy resolution 
of 15\%, as quoted by the EGRET collaboration in this energy range.
The photon spectrum is dominated by the sharp line coming from our
local halo, while the red-shifted extragalactic component is
somewhat fainter. 

If the gravitino is heavier than the $W$ or $Z$ bosons, new decay channels
are open. In this case, the energy spectrum consists of a 
continuous component, stemming from the fragmentation of the gauge bosons,
and a relatively intense gamma ray  line~\cite{Ibarra:2007wg}.

\section{Signatures at colliders}

In the scenario proposed in this work, the lifetime of the
NLSP is around one nanosecond or longer, giving rise to
very distinctive signatures at colliders.
When the NLSP is the lightest stau, that we assume mainly right-handed,
the main decay channel is 
${\widetilde \tau_R}\rightarrow \tau \;\nu_\mu, \mu\;\nu_\tau$. 
The corresponding decay length is
\begin{equation}
c \tau^{lep}_{\tilde\tau} \sim
30~{\rm cm}\left(  \frac{m_{\tilde{\tau}}}{200 {\rm GeV}} \right)^{-1}
\left(\frac{\lambda_{323}}{10^{-8}}\right)^{-2} \;.
\end{equation}
Besides, the small left-handed component of the 
stau mass eigenstate can trigger a decay into jets
through ${\widetilde \tau_L}\rightarrow b^c t$,
provided the process is kinematically open.
The hadronic decays are enhanced
compared to the leptonic decays by the larger bottom
Yukawa coupling and by the colour factor, but are
usually suppressed by the small left-right mixing.
The decay length in this channel reads
\begin{equation}
c \tau^{ had}_{\tilde\tau}\sim
8~{\rm m}\left(\frac{m_{\tilde{\tau}}}{200 {\rm GeV}} \right)^{-1}
\left(\frac{\lambda_{323}}{10^{-8}}\right)^{-2} 
\left(\frac{\cos\theta_\tau}{0.1}\right)^{-2}\;,
\end{equation}
where $\theta_\tau$ is the mixing angle of the staus.
Therefore, if the decay occurs inside the detector, the
signature would consist on a heavily ionizing charged
track followed by a muon track, by one jet or by three jets.

On the other hand, when the NLSP is the lightest neutralino, it
decays through $\chi^0_1\rightarrow \tau^{\pm} W^{\mp}$ if this decay
channel is kinematically accessible \cite{mrv98}.
The corresponding decay
length can be approximated by
\begin{equation}
c \tau_{\chi^0_1}\sim 1~{\rm m}
\left(\frac{m_{\chi^0_1}}{200~{\rm GeV}} \right)^{-3}
\left(\frac{\lambda_{323}}{10^{-8}}\right)^{-2} \;, 
\end{equation}
producing a significantly displaced vertex followed by jets,
that could be observed provided the decay occurs inside the
detector. If this decay channel is kinematically closed, 
the neutralino typically decays outside the detector yielding 
identical collider signatures to the case with $R$-parity conserved.

\section{Conclusions}

We have argued that  a supersymmetric scenario with gravitino LSP with 
a mass in the range 5--100 GeV and a small amount of $R$-parity violation,
$10^{-14}\lsim \lambda, \lambda' \lsim 10^{-7}$, yields a 
consistent thermal history of the Universe, in the sense that
it allows baryogenesis through thermal leptogenesis, 
provides a viable candidate for cold dark matter, and does
not spoil the successful predictions of the Standard Big Bang
Nucleosynthesis scenario.

We have presented a model that links $R$-parity breaking 
to $B-L$ breaking, and generates $R$-parity breaking parameters 
in the right range to accommodate the abovementioned scenario.
The relic gravitino decays into photons produce a diffuse halo 
and extragalactic gamma ray flux, that might have been observed
already by EGRET. Future experiments, such as GLAST or AMS-02
will provide unique opportunities to test the scenario of
decaying gravitino dark matter. Finally, we have discussed
the striking signatures that this scenario might produce
at future colliders, consisting in a vertex of the NLSP 
significantly displaced from the beam axis.

{\it Acknowledgments.} I would like to thank my collaborators G. Bertone, 
W. Buchm\"uller, L. Covi, K. Hamaguchi, D. Tran and
T. T. Yanagida for a very pleasant and fruitful collaboration.

\end{document}